\documentclass[11pt,showpacs,preprintnumbers,amsmath,amssymb,prd,nofootinbib,superscriptaddress]{revtex4-2}

\usepackage{dcolumn}
\usepackage{bm}
\usepackage{ifpdf}
\usepackage{hyperref}
\usepackage{bm}
\usepackage{xcolor,color,graphicx,graphics,physics}
\usepackage[spanish,english]{babel}
\usepackage[latin1]{inputenc}
\usepackage[OT1]{fontenc}
\usepackage{latexsym,amssymb,amsmath,amsfonts, slashed}
\usepackage{makeidx}
\usepackage{epsfig,subfigure}
\usepackage{natbib}
\usepackage{epstopdf}
\usepackage{mathrsfs}
\usepackage{hyperref}
\hypersetup{colorlinks=true, linkcolor=blue, citecolor=blue, urlcolor=blue}
\usepackage{enumerate}
\usepackage{cancel}

\usepackage{braket}
\usepackage{fixmath}

\everymath{\displaystyle}
\usepackage{graphicx}

\usepackage[T1]{fontenc}
\usepackage{amsmath}
\usepackage{amssymb}
\usepackage{graphicx}
\usepackage{xcolor}

\newcommand{\bea}{\begin{eqnarray}}
\newcommand{\eea}{\end{eqnarray}}

\newcommand{\orcid}[1]{\href{https://orcid.org/#1}{\includegraphics[width=10pt]{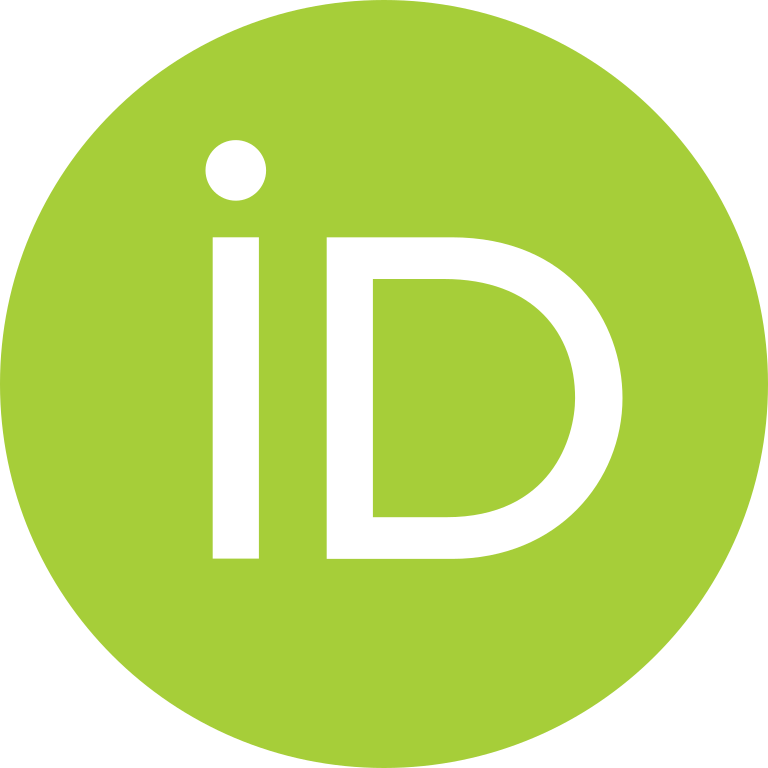}}}

\begin{document}

\title{ Scalar sector of the Myers and Pospelov model: thermal and size effects }

\author{L. H. A. R. Ferreira \orcid{0000-0002-4384-2545}}
\email{luiz.ferreira@fisica.ufmt.br }
\affiliation{Instituto de F\'{\i}sica, Universidade Federal de Mato Grosso,\\
78060-900, Cuiab\'{a}, Mato Grosso, Brazil}

\author{A. F. Santos \orcid{0000-0002-2505-5273}}
\email{alesandroferreira@fisica.ufmt.br}
\affiliation{Instituto de F\'{\i}sica, Universidade Federal de Mato Grosso,\\
78060-900, Cuiab\'{a}, Mato Grosso, Brazil}

\author{Carlos M. Reyes \orcid{0000-0001-5140-6658} }
\email{creyes@ubiobio.cl}
\affiliation{Centro de Ciencias Exactas,  
Universidad del B\'{\i}o-B\'{\i}o, \\ Casilla 447, Chill\'{a}n, Chile}

\begin{abstract}

The scalar sector of the Myers and Pospelov model is considered. This theory introduces a dimension 5 operator with a preferred four-vector which breaks 
Lorentz symmetry. We investigate various applications using the TFD formalism, a topological field theory that allows the study of thermal and size effects on an equal footing. In this context, Lorentz-violating corrections to the Casimir effect and Stefan-Boltzmann law have been calculated.

\end{abstract}

\maketitle

\section{Introduction}

Symmetry is a fundamental element in all systems of nature, and it emerges in different forms within a system. All physical theories are constructed under a set of symmetries. What does it mean, in physics, for a system to have some symmetry? It means it is invariant under a certain type of transformation.  As an example of a physical system constructed under a specific set of symmetries, let us consider the Standard Model (SM), a well-established theory describing all elementary particles and their interactions, except for the gravitational interaction, which is described by General Relativity (GR). The foundations of the SM are the Lorentz and CPT symmetries. Lorentz symmetry is a fundamental principle stating that experimental results are independent of the orientation or boost velocity of the laboratory through space. CPT symmetry predicts the equality of certain quantities, such as lifetime, mass, gyromagnetic ratio, and charge-to-mass ratio, for particles and antiparticles. Although the SM is a successful theory, it is not considered a fundamental theory because it excludes gravitational interaction and does not provide consistent answers for certain problems, such as dark matter, matter-antimatter asymmetry, the hierarchy problem, among others. In the last few decades, new and extended theories have been constructed to solve these problems. An important attempt is the development of models that search for a fundamental theory unifying the SM and GR. In these theories beyond the SM, tiny violations of Lorentz and CPT symmetries could emerge \cite{Kost1, Kost2}.

To study the violation of Lorentz and CPT symmetries in quantum field theories and gravity, a comprehensive framework has been constructed. This framework is called Standard Model Extension (SME) \cite{Kost3, Kost4}. This framework contains the known physics of the SM, GR, and all possible  operators that break Lorentz symmetry. It is important to emphasize that Lorentz symmetry is composed of two parts: observer Lorentz transformations and particle Lorentz transformations. A change of the frame is characterized by an observer Lorentz transformation, while a rotation or boost performed on an individual particle field or laboratory setting with a fixed coordinate frame consists of a particle Lorentz transformation. The SME is invariant under observer transformations.  Furthermore, the SME is divided into two sectors: (i) the minimal sector of the Standard Model Extension (mSME), which contains conventional quantization,  hermiticity, gauge invariance, power counting renormalizability, and positivity of the energy, and (ii) the nonminimal sector of the SME (nmSME), which is 
associated with operators of higher dimensions. In this work,  we study the scalar sector of the Myers and Pospelov model \cite{MP}. This model introduces 
a preferred four-vector $n$ that breaks Lorentz symmetry and couples to a five-dimensional operator. In the case the preferred vector has time components the theory develops higher-order time derivatives leading to the increase of degrees of freedom associated to ghost states. 
  
Several studies have been developed from this model. For example, the conservation of unitarity at one-loop order using the optical theorem and focusing on a quartic interaction term has been proved \cite{Reyes1}. Additionally, a perturbative expansion of the scalar sector in the Myers-Pospelov model, up to second order in the Lorentz-violating parameter and taking into account its higher-order time derivative character, has been constructed \cite{Pert}, among others. Here, the topological structure of the Thermo Field Dynamics (TFD) formalism is used to calculate corrections due to the Myers and Pospelov term for the Casimir effect and the Stefan-Boltzmann law associated with the massive scalar field.

The TFD formalism is a real-time thermal quantum field theory \cite{Umezawa1, Umezawa2, Umezawa22, quart, Khanna1, Khanna2} that was constructed by considering the statistical average of an arbitrary operator as the expectation value in a thermal vacuum. In order to construct the thermal vacuum, the Hilbert space is duplicated and the Bogoliubov transformation is introduced. TFD is a topological field theory that allows the investigation of different phenomena on an equal footing. Its topological structure is given as $\Gamma_D^d=(\mathbb{S}^1)^d\times \mathbb{R}^{D-d}$ with $1\leq d \leq D$. Here, $D$ represents the space-time dimensions and $d$ is the number of compactified dimensions. In this formalism, any set of dimensions of the manifold $\mathbb{R}^{D}$ can be compactified, where the circumference of the $nth$ $\mathbb{S}^1$ is specified by $\alpha_n$, which is the compactification parameter. Considering the scalar sector in the Myers-Pospelov model, three different topologies are investigated in this work. These applications involve introducing thermal and size effects into the theory on an equal footing. Two different phenomena, such as the Stefan-Boltzmann law and the Casimir effect, associated with the Myers-Pospelov model, are then studied. Consequently, the Lorentz-violating corrections to these effects are calculated and analyzed.

This paper is organized as follows. In Section II, the TFD formalism is briefly presented. In Section III, the scalar sector of the Myers-Pospelov model is introduced, and the energy-momentum tensor for this theory is calculated. To obtain a physical and finite quantity, some renormalization procedure is carried out. In Section IV, different topologies of the TFD structure are considered. In Subsections A and B, thermal and size effects are investigated separately. In Subsection C, both effects are analyzed, leading to the Casimir effect at finite temperature. In all applications, different directions of the four-vector $n^\mu$ are discussed. In Section V, some concluding remarks are provided.

\section{Thermo Field Dynamics Formalism}

In this section, a brief introduction to TFD formalism is presented. This formalism is constructed under two main ingredients. The first ingredient implies that the Hilbert space is doubled to create a thermal Hilbert space, defined as ${\cal S}_T={\cal S}\otimes \tilde{\cal S}$, where ${\cal S}$ represents the standard Hilbert space and $\tilde{\cal S}$ is the dual or tilde space.
There is a map between the tilde $\tilde{A_i}$ and non-tilde $A_i$ operators that is defined by the tilde (or dual) conjugation rules:
\bea
(A_iA_j)^\thicksim & =& \tilde{A_i}\tilde{A_j}, \quad (\tilde{A_i})^\thicksim = -\xi A_i,\\
(A_i^\dagger)^\thicksim &=& \tilde{A_i}^\dagger, \quad (cA_i+A_j)^\thicksim = c^*\tilde{A_i}+\tilde{A_j},\nonumber
\eea
where $\xi = -1\,(+1)$ for bosons (fermions). The second ingredient consists of the Bogoliubov transformation, which incorporates size and thermal effects through a rotation between tilde ($\tilde{\cal S}$) and non-tilde (${\cal S}$) spaces. As an example of the use of the Bogoliubov transformation, let's consider an arbitrary operator ${\cal O}$ in the Hilbert space ${\cal S}$ and its corresponding operator $\tilde{\cal O}$ in the tilde space $\tilde{\cal S}$. By applying this transformation, we obtain
\bea
\left( \begin{array}{cc} {\cal O}(k, \alpha)  \\\xi \tilde {\cal O}^\dagger(k, \alpha) \end{array} \right)={\cal U}(\alpha)\left( \begin{array}{cc} {\cal O}(k)  \\ \xi\tilde {\cal O}^\dagger(k) \end{array} \right),
\eea
where  ${\cal U}(\alpha)$ is the Bogoliubov transformation defined as
\bea
{\cal U}(\alpha)=\left( \begin{array}{cc} u(\alpha) & -w(\alpha) \\
\xi w(\alpha) & u(\alpha) \end{array} \right),
\eea
with $u^2(\alpha)+\xi w^2(\alpha)=1$. Here, $\alpha$ is the compactification parameter defined by $\alpha=(\alpha_0,\alpha_1,\cdots\alpha_{D-1})$, which is associated with the topological structure of the TFD formalism. Using the topological structure of this theory, any set of dimensions of the manifold can be compactified. For example, it is possible to choose a topology where $\alpha_0\equiv\beta$ and $\alpha_1,\cdots\alpha_{D-1}=0$, with $\beta=1/k_B T$ and $k_B$ being the Boltzmann constant. In this case, thermal effects are introduced into the theory. Alternatively, we can take $\alpha_3\neq 0$ and other components equal to zero. This introduces size effects. The main advantage is the fact that two different effects, thermal and size effects, can be implemented in the same way. 

Since our objective is to calculate some applications using the topological structure of TFD formalism for the scalar field with Lorentz violation, let's investigate how the Green function of the scalar field changes due to the compactification parameter $\alpha$. The propagator of the scalar field in the doubled notation of TFD formalism, with corrections due to the  $\alpha$ parameter  \cite{Umezawa2, quart}, is given as
\bea
G_0^{(AB)}(x-x';\alpha)=i\langle 0,\tilde{0}| \tau[\phi^A(x;\alpha)\phi^B(x';\alpha)]| 0,\tilde{0}\rangle,
\eea
where  $A$ and $B$ $=1,2$ and $\tau$ is the time ordering operator. To introduce the compactification parameter in the field, the Bogoliubov transformation is used, i.e.,
\bea
\phi(x;\alpha)&=&{\cal U}(\alpha)\phi(x){\cal U}^{-1}(\alpha).
\eea
In a similar way, the thermal vacuum is constructed as $|0(\alpha)\rangle={\cal U}(\alpha)|0,\tilde{0}\rangle$, with $|0,\tilde{0}\rangle=|0\rangle \otimes |\tilde{0}\rangle$. Note that $|0\rangle$ and $ |\tilde{0}\rangle$ are the vacuum states in the Hilbert and dual spaces, respectively. Then the propagator becomes
\bea
G_0^{(AB)}(x-x';\alpha)&=&i\langle 0(\alpha)| \tau[\phi^A(x)\phi^B(x')]| 0(\alpha)\rangle,\nonumber\\
&=&i\int \frac{d^4k}{(2\pi)^4}e^{-ik(x-x')}G_0^{(AB)}(k;\alpha),
\eea
where
\bea
G_0^{(AB)}(k;\alpha)={\cal U}^{-1}(\alpha)G_0^{(AB)}(k){\cal U}(\alpha),
\eea
with
\bea
G_0^{(AB)}(k)=\left( \begin{array}{cc} G_0(k) & 0 \\
0 & \xi G^*_0(k) \end{array} \right),
\eea
and $G_0(k)=(k^2-m^2+i\epsilon)^{-1}$, where $m$ is the scalar field mass. It is important to note that, the relevant quantities are associated with the non-tilde variables. Then the physical Green function is given by the component $A=B=1$, i.e.,
\bea
G_0^{(11)}(k;\alpha)=G_0(k)+\xi w^2(k;\alpha)[G^*_0(k)-G_0(k)],
\eea
where $w^2(k;\alpha)$ is given as
\bea
w^2(k;\alpha)=\sum_{s=1}^d\sum_{\lbrace\sigma_s\rbrace}2^{s-1}\sum_{l_{\sigma_1},...,l_{\sigma_s}=1}^\infty(-\xi)^{s+\sum_{r=1}^sl_{\sigma_r}}\,\exp\left[{-\sum_{j=1}^s\alpha_{\sigma_j} l_{\sigma_j} k^{\sigma_j}}\right].\label{BT}
\eea
This is the generalized Bogoliubov transformation \cite{GBT}. Here $d$ is the number of compactified dimensions, $\lbrace\sigma_s\rbrace$ denotes the set of all combinations with $s$ elements and $k$ is the 4-momentum.

In the next section, the Lorentz-violating extension of the Myers-Pospelov model is described. The energy-momentum tensor for this theory, considering the TFD formalism, is calculated.

\section{The model and the energy-momentum tensor} 

The Lagrangian describing the Lorentz violating extension Myers-Pospelov in the scalar sector is given as
\bea 
{\cal L}&=&\partial_\mu\phi^\dagger\partial^\mu\phi-m^2\phi^\dagger\phi+ig\phi^\dagger(n\cdot\partial)^3\phi,
\eea
where the third term encodes the breakdown of Lorentz symmetry, the constant vector $n^\mu$ acts as a background field and $g$ is the coupling constant. It is convenient to use the symmetrized Lagrangian
\bea 
{\cal L}&=&\partial_\mu\phi^\dagger\partial^\mu\phi-m^2\phi^\dagger\phi+\frac{ig}{2}\left(\phi^\dagger(n\cdot\partial)^3\phi-((n\cdot\partial)^3\phi^\dagger)\phi\right).\label{2}
\eea

To find the equation of motion we use the extended Euler-Lagrange equation, i.e.,
\bea
0=\frac{\partial{\cal L}}{\partial \phi}-\partial_\mu\frac{\partial{\cal L}}{\partial(\partial_\mu\phi)}-\partial_{\nu 1}\partial_{\nu 2}\partial_{\mu}\left(\frac{\partial{\cal L}}{\partial({\partial_{\nu 1}\partial_{\nu 2}\partial_{\mu}\phi})}\right).
\eea
We have
\bea
\frac{\partial{\cal L}}{\partial \phi}&=&-m^2\phi^\dagger-\frac{ig}{2}(\partial\cdot n)^3\phi^\dagger,\\
\frac{\partial{\cal L}}{\partial(\partial_\mu\phi)}&=&\partial^\mu\phi^\dagger,\\
\frac{\partial{\cal L}}{\partial({\partial_{\nu 1}\partial_{\nu 2}\partial_{\mu}\phi})}&=&\frac{ig}{2}\phi^\dagger n^{\nu1}n^{\nu2}n^{\mu}.
\eea
Then the equation of motion becomes
\bea
-m^2\phi^\dagger-\Box\phi^\dagger-ig(\partial \cdot n)^3\phi^\dagger=0.
\eea
Analogously, by taking the dagger operation we obtain the equation of motion for the field $\phi$
\bea
-m^2\phi-\Box\phi+ig(\partial \cdot n)^3\phi=0.
\eea

In order to investigate some applications using TFD formalism the main quantity that must be calculated is the energy-momentum tensor associated with the theory. In the ausence of second-order derivatives, we use the expression for the energy-momentum tensor \cite{Moeller}
\bea
T^{\mu}_\alpha&=&-\delta^\mu_\alpha {\cal L}+\frac{\partial{\cal L}}{\partial(\partial_\mu\phi)}\partial_\alpha\phi+\partial_\alpha\phi^\dagger\frac{\partial{\cal L}}{\partial(\partial_\mu\phi^\dagger)}+\Biggl\{\partial_{\nu1}\partial_{\nu2}\left(\frac{\partial{\cal L}}{\partial(\partial_{\nu1}\partial_{\nu2}\partial_\mu\phi)}\right)\partial_\alpha\phi\nonumber\\
&-&\partial_{\nu1}\left(\frac{\partial{\cal L}}{\partial(\partial_{\nu1}\partial_{\nu2}\partial_\mu\phi)}\right)\partial_{\nu2}\partial_\alpha\phi+\left(\frac{\partial{\cal L}}{\partial(\partial_{\nu1}\partial_{\nu2}\partial_\mu\phi)}\right)\partial_{\nu1}\partial_{\nu2}\partial_\alpha\phi\Biggl\}\nonumber\\
&+&\Biggl\{(\partial_\alpha\phi^\dagger)\partial_{\nu1}\partial_{\nu2}\left(\frac{\partial{\cal L}}{\partial(\partial_{\nu1}\partial_{\nu2}\partial_\mu\phi)}\right)-(\partial_{\nu2}\partial_\alpha\phi^\dagger)\partial_{\nu1}\left(\frac{\partial{\cal L}}{\partial(\partial_{\nu1}\partial_{\nu2}\partial_\mu\phi)}\right)\nonumber\\
&+&\partial_{\nu1}\partial_{\nu2}\partial_\alpha\phi^\dagger\left(\frac{\partial{\cal L}}{\partial(\partial_{\nu1}\partial_{\nu2}\partial_\mu\phi)}\right)
\Biggl\}.
\eea
Let us write each piece
\bea
\frac{\partial{\cal L}}{\partial(\partial_\mu\phi)}=\partial^\mu\phi^\dagger, \quad\quad\quad \frac{\partial{\cal L}}{\partial(\partial_\mu\phi^\dagger)}=\partial^\mu\phi,\label{10}
\eea
\bea
\frac{\partial{\cal L}}{\partial(\partial_{\nu1}\partial_{\nu2}\partial_\mu\phi)}=\frac{ig}{2}\phi^\dagger n^{\nu1}n^{\nu2}n^{\mu},  \quad\quad\quad \frac{\partial{\cal L}}{\partial(\partial_{\nu1}\partial_{\nu2}\partial_\mu\phi^\dagger)}=-\frac{ig}{2}\phi n^{\nu1}n^{\nu2}n^{\mu},\label{11}
\eea
\bea
\partial_{\nu1}\partial_{\nu2}\frac{\partial{\cal L}}{\partial(\partial_{\nu1}\partial_{\nu2}\partial_\mu\phi)}=\frac{ig}{2}n^{\mu}(n\cdot \partial)^2\phi^\dagger,  \quad\quad\quad \partial_{\nu1}\partial_{\nu2}\frac{\partial{\cal L}}{\partial(\partial_{\nu1}\partial_{\nu2}\partial_\mu\phi^\dagger)}=-\frac{ig}{2}n^{\mu}(n\cdot \partial)^2\phi,\label{12}
\eea
\bea
\partial_{\nu1}\frac{\partial{\cal L}}{\partial(\partial_{\nu1}\partial_{\nu2}\partial_\mu\phi)}=\frac{ig}{2}n^{\mu}n^{\nu2}(n\cdot \partial)\phi^\dagger,  \quad\quad\quad \partial_{\nu1}\frac{\partial{\cal L}}{\partial(\partial_{\nu1}\partial_{\nu2}\partial_\mu\phi^\dagger)}=-\frac{ig}{2}n^{\mu}n^{\nu2}(n\cdot \partial)\phi.\label{13}
\eea

Using these pieces, Eqs. (\ref{10}), (\ref{11}), (\ref{12}) and (\ref{13}), and the Lagrangian (\ref{2}), the energy-momentum tensor is given as
\bea
T^{\mu\nu}&=&-\eta^{\mu\nu}\left(\partial^\lambda\phi^\dagger\partial_\lambda\phi-m^2\phi^\dagger\phi\right)+\partial^\mu\phi^\dagger\partial^\nu\phi+\partial^\nu\phi^\dagger\partial^\mu\phi\nonumber\\
&+&\frac{ig}{2}\Bigl[-\eta^{\mu\nu}\left(\phi^\dagger(n\cdot\partial)^3\phi-((n\cdot\partial)^3\phi^\dagger)\phi\right)\nonumber\\
&+&n^\mu\Bigl((n\cdot\partial)^2\phi^\dagger\partial^\nu\phi-(n\cdot\partial)\phi^\dagger(n\cdot\partial)\partial^\nu\phi+\phi^\dagger(n\cdot\partial)^2\partial^\nu\phi\nonumber\\
&-&\partial^\nu\phi^\dagger(n\cdot\partial)^2\phi+(n\cdot\partial)\partial^\nu\phi^\dagger(n\cdot\partial)\phi- (n\cdot\partial)^2(\partial^\nu\phi^\dagger)\phi
\Bigl)
\Bigl].
\eea

To simplify, let us rewrite the energy-momentum tensor as
\bea
T^{\mu\nu}=T^{\mu\nu}_{LI}+T^{\mu\nu}_{LV},\label{15}
\eea
where the Lorentz invariant part is given as
\bea
T^{\mu\nu}_{LI}=-\eta^{\mu\nu}\left(\partial^\lambda\phi^\dagger\partial_\lambda\phi-m^2\phi^\dagger\phi\right)+2\partial^\mu\phi^\dagger\partial^\nu\phi
\eea
 and the Lorentz violating part is
\bea
T^{\mu\nu}_{LV}&=&\frac{ig}{2}\Bigl[-\eta^{\mu\nu}\left(\phi^\dagger(n\cdot\partial)^3\phi-((n\cdot\partial)^3\phi^\dagger)\phi\right)\nonumber\\
&+&n^\mu\Bigl((n\cdot\partial)^2\phi^\dagger\partial^\nu\phi-(n\cdot\partial)\phi^\dagger(n\cdot\partial)\partial^\nu\phi+\phi^\dagger(n\cdot\partial)^2\partial^\nu\phi\nonumber\\
&-&\partial^\nu\phi^\dagger(n\cdot\partial)^2\phi+(n\cdot\partial)\partial^\nu\phi^\dagger(n\cdot\partial)\phi- (n\cdot\partial)^2(\partial^\nu\phi^\dagger)\phi
\Bigl)
\Bigl].
 \eea

In order to avoid divergences, the energy-momentum tensor is written at different points in space-time, i.e.,
\bea
T^{\mu\nu}_{LI}(x)&=&\lim_{x'\to x}\tau\Bigl\{-\eta^{\mu\nu}\left(\partial^\lambda\phi(x)^\dagger\partial'_\lambda\phi(x')-m^2\phi(x)^\dagger\phi(x')\right)+2\partial^\mu\phi(x)^\dagger\partial'^\nu\phi(x')\Bigl\},\label{EMT1}
\eea
where $\tau$ is the ordering operator. In the same form the Lorentz violating part is written as
\bea
T^{\mu\nu}_{LV}&=&\lim_{x'\to x}\tau\Biggl\{\frac{ig}{2}\Bigl[-\eta^{\mu\nu}\left(\phi(x)^\dagger(n\cdot\partial')^3\phi(x')-((n\cdot\partial)^3\phi(x)^\dagger)\phi(x')\right)\nonumber\\
&+&n^\mu\Bigl((n\cdot\partial)^2\phi(x)^\dagger\partial'^\nu\phi(x')-(n\cdot\partial)\phi(x)^\dagger(n\cdot\partial')\partial'^\nu\phi(x')+\phi(x)^\dagger(n\cdot\partial')^2\partial'^\nu\phi(x')\nonumber\\
&-&\partial^\nu\phi(x)^\dagger(n\cdot\partial')^2\phi(x')+(n\cdot\partial)\partial^\nu\phi(x)^\dagger(n\cdot\partial')\phi(x')- (n\cdot\partial)^2(\partial^\nu\phi(x)^\dagger)\phi(x')
\Bigl)
\Bigl]\Biggl\}.\label{EMT2}
 \eea

To proceed with some applications, let's use the time ordering operator $\tau$ and consider the commutation relations in higher order time derivatives theories. 

In principle we have a higher-order theory and for some choices of the preferred four-vector $n^{\mu}$ we have an increase in the degrees of freedom. This means 
that phase space is increased since we also have a momentum variable associated to velocities or time derivative of the fields $\dot \phi$. To deal with this issue and control the increase of degrees of freedom, here
we apply the perturbative method developed in~\cite{Cheng:2002rz}, which by working at the level of the symplectic structure of the theory,
allows to preserve perturbative degrees of freedom, suppress the ghost-states modes and produce a positive-definite Hamiltonian. The perturbative reduction was
 implemented for the scalar Myers-Pospelov model
 through a
second order expansion in $g$ of the symplectic two form $\Omega$ given in~\cite{Pert}.
In this context, the quantization of the Myers-Pospelov (MP) model provides perturbative corrections to the standard commutation relations at equal time \cite{Pert}, i. e., 
\bea 
[\phi(x),\phi^\dagger(x')]&=&-ig\delta^3(\vec{x} - \vec{x'}),
\eea
\bea
[\phi(x),\partial'^\mu\phi^\dagger(x')]&=&v_0^\mu(1+3g^2E^2)\delta^3({\vec{x}-\vec{x'}}),
\eea
\bea
[\phi^\dagger(x),\partial'^\mu\phi(x')]&=&v_0^\mu(1+3g^2E^2)\delta^3({\vec{x}-\vec{x'}}),
\eea
\bea 
[\partial^\mu\phi(x),\partial'^\mu\phi^\dagger(x')]&=&-2iv_0^\mu g E^2\delta^3({\vec{x}-\vec{x'}}).
\eea
where $v_0^\mu=(1,0,0,0)$ is a time-like vector, $\partial^\mu\theta(x_0-x_0')=v_0^\mu\delta(x_0-x_0')$, $g$ is the coupling constant and $E^2=m^2-\nabla_x^2$. Using these ingredients, Eq. (\ref{EMT1}) can written as
\bea
T^{\mu\nu}_{LI}(x)=\lim_{x'\to x}\Bigl\{\Pi^{\mu\nu}\tau[\phi^\dagger(x)\phi(x')]+\Theta^{\mu\nu}\delta(x-x')\Bigg\}
\eea
with 
\bea
\Pi^{\mu\nu}&=&-\eta^{\mu\nu}\left(\partial^\lambda\partial'_\lambda-m^2\right)+2\partial^\mu\partial'^\nu,\\
\Theta^{\mu\nu}&=&2(1+3g^2E^2)(\eta^{\mu\nu}+v_0^\mu v_0^\nu).
\eea
In a similar form, after some calculations, Eq. (\ref{EMT2}) becomes
\bea
T^{\mu\nu}_{LV}(x)=\frac{ig}{2}\lim_{x'\to x}\Bigl\{\Gamma^{\mu\nu}\tau[\phi^\dagger(x)\phi(x')]+I^{\mu\nu}(x-x')\Bigg\}
\eea
where
\bea
\Gamma^{\mu\nu}&=&-\eta^{\mu\nu}\left((n\cdot\partial')^3-((n\cdot\partial)^3\right)\nonumber\\
&+&n^\mu\Bigl((n\cdot\partial)^2\partial'^\nu-(n\cdot\partial)(n\cdot\partial')\partial'^\nu+(n\cdot\partial')^2\partial'^\nu\nonumber\\
&-&\partial^\nu(n\cdot\partial')^2+(n\cdot\partial)\partial^\nu(n\cdot\partial')- (n\cdot\partial)^2\partial^\nu\Bigl),\label{25}
\eea
and
\bea
I^{\mu\nu}(x-x')&=&-ig\eta^{\mu\nu}\delta^3(\vec{x}-\vec{x'})\left[(n\cdot\partial')^3-(n\cdot\partial)^3\right]\theta(x_0-x_0')\nonumber\\
&+&n^\mu(1+3g^2E^2)\Bigl\{\delta^3(\vec{x}-\vec{x'})\Bigl[v_0^\mu(n\cdot\partial)^2+n_\rho v_0^\rho(n\cdot\partial')\partial'^\nu\Bigl]\nonumber\\
&-&(n\cdot\partial')v_0^\nu\delta^3(\vec{x}-\vec{x'})(n\cdot\partial)-(n\cdot\partial')^2v_0^\nu\delta^3(\vec{x}-\vec{x'})\nonumber\\
&-&(n\cdot\partial)v_0^\nu\delta^3(\vec{x}-\vec{x'})(n\cdot\partial')+n_\rho v_0^\rho\delta^3(\vec{x}-\vec{x'})(n\cdot\partial)\partial^\nu\Bigl\}\theta(x_0-x_0')\nonumber\\
&+&ign^\mu\Bigl[(n\cdot\partial)^2\delta^3(\vec{x}-\vec{x'})v_0^\nu \delta(x_0-x_0')+\delta^3(\vec{x}-\vec{x'})(n\cdot\partial')^2\partial'^\nu \theta(x_0-x_0')\nonumber\\
&-&\partial^\nu\delta^3(\vec{x}-\vec{x'})(n\cdot\partial')^2\theta(x_0-x_0')-\delta^3(\vec{x}-\vec{x'})(n\cdot\partial)^2\partial^\nu \theta(x_0-x_0')\Bigl].
\eea

Taking the vacuum expectation value of Eq. (\ref{15}), we get
\bea
\braket{T^{\mu\nu}(x)}&\equiv&\braket{0|T^{\mu\nu}(x)|0}\nonumber\\
	&=&\lim_{x'\to x}{\bigg\{}\Pi^{\mu\nu}\braket{0|\tau[\phi^\dagger(x)\phi(x')]|0}+\Theta^{\mu\nu}\delta(x-x')\braket{0|0}\nonumber\\
	&+&\frac{ig}{2}\left(\Gamma^{\mu\nu}\braket{0|\tau[\phi^\dagger(x)\phi(x')]|0}+I^{\mu\nu}(x,x')\braket{0|0}\right){\bigg\}}.\label{VEV}
\eea
Using the definition of the propagator for a massive scalar field, i.e.,
\bea
\braket{0|\tau[\phi^\dagger(x)\phi(x')]|0}&=iG_0(x-x')
\eea
where
\bea
G_0(x-x')=-\frac{im}{4\pi^2}\frac{K_1(m\sqrt{-(x-x')^2})}{\sqrt{-(x-x')^2}}
\eea
with $ K_\nu(z) $ being the Bessel function, Eq. (\ref{VEV}) becomes
\bea
\braket{T^{\mu\nu}(x)}&=&\lim_{x'\to x}{\bigg\{}i\Pi^{\mu\nu}G_0(x-x')+\Theta^{\mu\nu}\delta(x-x')\nonumber\\
	&+&\frac{ig}{2}\left(i\Gamma^{\mu\nu}G_0(x-x')+I^{\mu\nu}(x,x')\right){\bigg\}}.
\eea

Considering the topological structure of the TFD approach, this equation is written
\bea
\braket{T^{\mu\nu}(x;\alpha)}&=&\lim_{x'\to x}{\bigg\{}i\Pi^{\mu\nu}G_0^{(ab)}(x-x';\alpha)+\Theta^{\mu\nu}\delta(x-x')\nonumber\\
	&+&\frac{ig}{2}\left(i\Gamma^{\mu\nu}G_0^{(ab)}(x-x';\alpha)+I^{\mu\nu}(x,x')\right){\bigg\}}.
\eea
where $a, b = 1, 2$ denoting the doubled notation.

To obtain a physical or finite energy-momentum tensor, the Casimir prescription is used. Then
\bea
{\cal T}^{\mu\nu(ab)}(x;\alpha)&=&\braket{T^{\mu\nu(ab)}(x;\alpha)}-\braket{T^{\mu\nu(ab)}(x)}\nonumber\\
	&=&\lim_{x'\to x}{\bigg\{}\left[i\Pi^{\mu\nu}-\frac{g}{2}\Gamma^{\mu\nu}\right]\overline{G}^{(ab)}_0(x-x';\alpha){\bigg\}},\label{EMT3}
\eea
with
\bea
\overline{G}^{(ab)}_0(x-x';\alpha)=G_0^{(ab)}(x-x';\alpha)-G^{(ab)}_0(x-x').
\eea

In the next section,  Eq. (\ref{EMT3}) is used to investigate thermal and size effects for different choices of the constant vector $ n^\mu $, which lead to the breaking of Lorentz symmetries.

\section{Thermal and size Applications}

To investigate the thermal and size effects associated with a massive scalar field with higher-order Lorentz symmetry breaking, let us consider the topological structure of the TFD formalism. Using this approach it is possible to analyze different phenomena on an equal footing. Here three different topologies are considered: 1 - The topology $\Gamma_4^1=\mathbb{S}^1\times\mathbb{R}^{3}$, where $\alpha=(\beta,0,0,0)$. Here, the time-axis is compactified in $\mathbb{S}^1$, with circumference $\beta$. 2 - The topology $\Gamma_4^1$ with $\alpha=(0,0,0,i2d)$, where the compactification along the coordinate $z$ is considered. 3 - The topology $\Gamma_4^2=\mathbb{S}^1\times\mathbb{S}^1\times\mathbb{R}^{2}$ with $\alpha=(\beta,0,0,i2d)$ is used. In this case, two axes are compactified, time and coordinate $z$.

\subsection{Thermal effects}

To analyze thermal effects the compactification parameter is taken as $\alpha=(\beta,0,0,0)$. This compactification allows us to calculate the Stefan-Boltzmann law associated with the Lorentz violation theory. For such a development, the Bogoliubov transformation is given as
\bea
\omega^2(\beta)=\sum^{\infty}_{l_0=1}e^{-\beta k^0l_0},
\eea
and the Green function is
\bea
\overline{G}_0(x-x';\beta)=2\sum^{\infty}_{l_0=1}G_0(x-x'-i\beta l_0v_0),
\eea
where $ v^\mu_0=(1,0,0,0) $.

Now let us analyze the energy-momentum tensor given in Eq. (\ref{EMT3}) for different choices of the constant vector $n^\mu$ that acts as a background field.

\subsubsection{Time-like constant vector}

In this case, the constant vector is chosen as $ n^\mu=(1,0,0,0) $. Then, the energy-momentum tensor for $\mu=\nu=0$ becomes
\bea
{\cal T}^{00(11)}(x;\alpha)=2\lim_{x'\to x}\sum_{l_0=1}^{\infty}{\bigg\{}\left[i\Pi^{00}-\frac{g}{2}\Gamma^{00}\right]G_0(x-x'-i\beta l_0v_0){\bigg\}},
\eea
where
\bea
\Pi^{00}&=&\partial^0\partial'^0+\partial^1\partial'^1+\partial^2\partial'^2+\partial^3\partial'^3+m^2,\label{37}\\
\Gamma^{00}&=&2\partial^0\partial^0\partial'^0-2\partial^0\partial'^0\partial'^0.\label{38}
\eea
This expression leads to the Stefan-Boltzmann law associated with the massive scalar field, i.e.,
\bea
{\cal T}^{00(11)}(\beta)&=&\frac{m^2}{\pi^2\beta^2}\sum^{\infty}_{l_0=1}\frac{1}{l_0^2}\Biggl\{(m\beta l_0)K_1(m\beta l_0)+3K_2(m\beta l_0)\nonumber\\
&+&\frac{g}{\beta l_0}\left[(m\beta l_0)^2K_0(m\beta l_0)+5(m\beta l_0)K_1(m\beta l_0)+12 K_2(m\beta l_0)\right]\Biggl\}.\label{39}
\eea
It is important to note that for this choice of vector $n^\mu$ the Lorentz violating term contributes to the energy associated with the scalar field. 

To be clearer about the correction due to the higher-order Lorentz symmetry breaking, let us consider the limit of small mass bound, i.e., $ m\to 0 $. Then Eq. (\ref{39}) becomes
\bea
{\cal T}^{00(11)}(\beta)&=\frac{\pi^2}{15\beta^4}+\frac{24g\zeta(5)}{\pi^2\beta^5},\label{Pressure}
\eea
where $\zeta(x)$ is the Riemann zeta function. Here has been used that in such a limit the Bessel function is given as
\bea
K_\nu(z)\approx\Gamma(\nu)2^{\nu-1}z^{-\nu}.
\eea
It is important to note in Eq. (\ref{Pressure}) that the correction for Lorentz violation contributes to an increase in energy. Furthermore, the temperature dependence of the Lorentz-violating component does not follow $T^4$, but rather $T^5$, indicating that at high temperatures, this term can become dominant.

\subsubsection{Space-like constant vector}

Here, the  constant vector $n^\mu$ is chosen as $ n^\mu=(0,1,0,0) $, $ n^\mu=(0,0,1,0) $ and $ n^\mu=(0,0,0,1) $. Following steps similar to the previous subsection,
it is noted that the higher-order Lorentz violation term does not contribute to these choices of $n^\mu$. Then the Stefan-Boltzmann law is the standard result associated to the massive scalar field, i.e.,
\bea
{\cal T}^{00(11)}(\beta)&=&\frac{m^2}{\pi^2\beta^2}\sum^{\infty}_{l_0=1}\frac{1}{l_0^2}\Biggl\{(m\beta l_0)K_1(m\beta l_0)+3K_2(m\beta l_0)]\Biggl\}.
\eea

Therefore, the correction due to Lorentz violation depends on the direction of the constant vector that acts as a background field. 

\subsection{Size effects}

In this section, to investigate the size effects, the compactification parameter is taken as $\alpha=(0,0,0,i2d)$. This means that the compactification along the coordinate $z$ is considered, which allows calculating the Casimir effect at zero temperature associated with the massive scalar field with higher-order Lorentz symmetry breaking.

For this topology, the Bogoliubov transformation is given as
\bea
\omega^2(d)=\sum^{\infty}_{l_3=1}e^{-i2dk^3l_3},
\eea
and the Green functions is written as
\bea
\overline{G}_0(x-x';d)=2\sum^{\infty}_{l_3=1}G_0(x-x'-2dl_3v_3),
\eea
with $ v^\mu_3=(0,0,0,1) $. In this context, the Casimir effect at zero temperature will be calculated for different direction of the constant vector $n^\mu$.

\subsubsection{Time-like constant vector}

Assuming $ n^\mu=(1,0,0,0)$ the component $ \mu=\nu=0 $ of the energy-momentum tensor becomes
\bea
{\cal T}^{00(11)}(x;d)=2\lim_{x'\to x}\sum_{l_3=1}^{\infty}{\bigg\{}\left[i\Pi^{00}-\frac{g}{2}\Gamma^{00}\right]G_0(x-x'-2dl_3v_3){\bigg\}},
\eea
using Eqs. (\ref{37}) and (\ref{38}), it is obtained that
\bea
{\cal T}^{00(11)}(d)=-\frac{m^2}{4\pi^2 d^2}\sum_{l_3=1}^{\infty}\frac{K_2(2mdl_3)}{l_3^2}.\label{46}
\eea
This is the Casimir energy for the massive scalar field, where we can see that the violation of Lorentz symmetries has no influence on its value. In the following let us calculate the Casimir pressure which is given by
\bea
{\cal T}^{33(11)}(x;d)=2\lim_{x'\to x}\sum_{l_3=1}^{\infty}{\bigg\{}\left[i\Pi^{33}-\frac{g}{2}\Gamma^{33}\right]G_0(x-x'-2dl_3v_3){\bigg\}},
\eea
where
\bea
\Pi^{33}&=&\partial^0\partial'^0-\partial^1\partial'^1-\partial^2\partial'^2+\partial^3\partial'^3-m^2,\\
\Gamma^{33}&=&\partial'^0\partial'^0\partial'^0-\partial^0\partial^0\partial^0.
\eea
After some calculations we find
\bea
{\cal T}^{33(11)}(d)=-\frac{m^2}{4\pi^2d^2}\sum_{l_3=1}^{\infty}\frac{1}{l^2_3}{\bigg\{}2mdl_3K_1(2mdl_3)+3K_2(2mdk_3){\bigg\}}.\label{50}
\eea
Similar to what happens in the Casimir energy, in the Casimir pressure there is no contribution due to the Lorentz violation when the constant vector $n^\mu$ is a time-like vector. Let us then analyze the cases in which the constant vector has another direction.

\subsubsection{Space-like constant vector}

Although there are three possibilities for the case where the vector $n^\mu$ is a space-like constant vector, two of them $ n^\mu=(0,1,0,0) $ and $ n^\mu=(0,0,1,0) $ displays the same result, Eqs. (\ref{46}) and (\ref{50}), obtained for the time-like case. Therefore, special attention should be paid to the case $ n^\mu=(0,0,0,1) $. 

The last case, which chooses the space-like constant vector as $ n^\mu=(0,0,0,1) $, leads to an imaginary Casimir energy. The same happens with Casimir pressure. However,  this application leads to the Casimir effect, a physical and measurable phenomenon. Therefore, it makes no sense to obtain an imaginary value for this quantity.

In order to derive real and physical corrections resulting from Lorentz violation for the Casimir effect, the space-like constant vector must be chosen as $ n^\mu=(0,0,0,i) $. For this choice, the components $ \mu=\nu=0 $  and $ \mu=\nu=3 $ of Eq. (\ref{25}) are given as
\bea
\Gamma^{00}&=&-i(\partial'^3\partial'^3\partial'^3-\partial^3\partial^3\partial^3),\\
\Gamma^{33}&=&-2i(\partial^3\partial^3\partial'^3-\partial^3\partial'^3\partial'^3).
\eea

With these ingredients the Casimir energy with Lorentz violation is found as
\bea
{\cal T}^{00(11)}(d)&=&-\frac{m^2}{4\pi^2 d^2}\sum_{l_3=1}^{\infty}\frac{1}{l_3^2}\Bigl\{K_2(2mdl_3)\nonumber\\
&+&\frac{g}{2l_3}\left[2(dml_3)^2K_0(2mdl_3)+5(dml_3)K_1(2mdl_3)+6K_2(2mdl_3)\right]\Bigl\}
\eea
and the Casimir pressure is
\bea
{\cal T}^{33(11)}(d)&=&-\frac{m^2}{4\pi^2 d^2}\sum_{l_3=1}^{\infty}\frac{1}{l_3^2}\Bigl\{2mdl_3 K_1(2mdl_3)+3K_2(2mdl_3)\nonumber\\
&+&\frac{g}{l_3}\left[2(dml_3)^2K_0(2mdl_3)+5(dml_3)K_1(2mdl_3)+6K_2(2mdl_3)\right]\Bigl\}\label{54}
\eea
It's important to note that the corrections arising from Lorentz violation contribute to an increase in both the Casimir energy and pressure.

Looking at the Casimir force, expression (\ref{54}), in the small mass limit, we get
\bea
{\cal T}^{00(11)}=-\frac{\pi^2}{240d^4}-\frac{3g\zeta(5)}{4\pi^2d^5}.\label{Eii}
\eea
This result demonstrates that the Lorentz violation correction exhibits a different dependence on $d$, although it remains an attractive contribution, similar to the usual case.

It is important to observe that there is no contribution to the Stefan-Boltzmann law due to this case with $ n^\mu=(0,0,0,i) $.

\subsection{Thermal and Size effects}

In this subsection, two compactifications are considered, in the time and in the $z$ direction. As a consequence, thermal and size effects are analyzed together, i.e., the Casimir effect at finite temperature to the massive scalar field with higher-order Lorentz symmetry breaking is calculated. For this application the compactification parameter is assumed as $\alpha=(\beta,0,0,i2d)$. Then, the Bogoliubov transformation is written as
\bea
\omega^2(\beta,d)=\sum_{l_0=1}^{\infty}e^{-\beta k^0l_0}+\sum_{l_3=1}^{\infty}e^{-i2dk^3l_3}+2\sum_{l_0,l_3=1}^{\infty}e^{-\beta k^0l_0-i2dk^3l_3},
\eea
where these terms are associated with the thermal effect, size effect and thermal and size effects together, respectively. As the interest here is the combined effects, given by the third term, the Green function associated with it is given as
\bea
\overline{G}_0(x-x';\beta,d)=4\sum_{l_0,l_3=1}^{\infty}G_0(x-x'-i\beta v_0l_0-2dl_3v_3).
\eea

Now the energy-momentum tensor is calculated for different choices of the constant vector $ n^\mu $.

\subsubsection{Time-like constant vector}

Considering $ n^\mu=(1,0,0,0) $, the total energy-momentum tensor with $ \mu=\nu=0 $ becomes
\bea
{\cal T}^{00(11)}(\beta,d)&=&{\cal T}^{00(11)}_{LI}(\beta,d)+{\cal T}^{00(11)}_{LV}(\beta,d).
\eea
This give us the Casimir energy at finite temperature with corrections due to Lorentz violation, where
{\small
\bea
{\cal T}^{00(11)}_{LI}(\beta,d)&=&-\frac{2m}{\pi^2}\sum_{l_0,l_3=1}^{\infty}\Biggl\{\frac{m\left((2dl_3)^2-3(\beta l_0)^2\right)}{[(2dl_3)^2+(\beta l_0)^2]^{2}}K_0\left(z\right)\nonumber\\
&+&\frac{\left(2(2dl_3)^2-2(\beta l_0)^2(3+2(dml_3)^2)-m^2(\beta l_0)^4\right)}{[(2dl_3)^2+(\beta l_0)^2]^{5/2}}K_1\left(z\right)
\Biggl\}\label{LI00}
\eea
and 
\bea
{\cal T}^{00(11)}_{LV}(\beta,d)&=&\frac{2\beta g m}{\pi^2}\sum_{l_0,l_3=1}^{\infty}l_0\Biggl\{\frac{m\left((2\beta l_0)^2((dml_3)^2+3)-12(2dl_3)^2+m^2(\beta l_0)^4\right)}{[(2dl_3)^2+(\beta l_0)^2]^{3}}K_0\left(z\right)\nonumber\\
&-&\frac{\left(-8(\beta l_0)^2((dml_3)^2+3)+12(2dl_3)^2((dml_3)^2+2)-5m^2(\beta l_0)^4\right)}{[(2dl_3)^2+(\beta l_0)^2]^{7/2}}K_1\left(z\right)
\Biggl\}
\eea
}
with $z=m\sqrt{(2dl_3)^2+(\beta l_0)^2}$. Considering $ \mu=\nu=3$ and following the same steps we get the Casimir pressure at finite temperature
\bea
{\cal T}^{33(11)}(\beta,d)&=&{\cal T}^{33(11)}_{LI}(\beta,d)+{\cal T}^{33(11)}_{LV}(\beta,d),
\eea
where
{\small
\bea
{\cal T}^{33(11)}_{LI}(\beta,d)&=&-\frac{2m}{\pi^2}\sum_{l_0,l_3=1}^{\infty}\Biggl\{\frac{m\left(3(2dl_3)^2-(\beta l_0)^2\right)}{[(2dl_3)^2+(\beta l_0)^2]^{2}}K_0\left(z\right)\nonumber\\
&+&\frac{2\left(-(\beta l_0)^2+2(dl_3)^2(6+m^2(dml_3)^2+(\beta l_0)^2)\right)}{[(2dl_3)^2+(\beta l_0)^2]^{5/2}}K_1\left(z\right)
\Biggl\}\label{LI33}
\eea
and 
\bea
{\cal T}^{33(11)}_{LV}(\beta,d)&=&\frac{\beta g m}{\pi^2}\sum_{l_0,l_3=1}^{\infty}l_0\Biggl\{\frac{m\left((2\beta l_0)^2((dml_3)^2+3)-12(2dl_3)^2+m^2(\beta l_0)^4\right)}{[(2dl_3)^2+(\beta l_0)^2]^{3}}K_0\left(z\right)\nonumber\\
&-&\frac{\left(-8(\beta l_0)^2((dml_3)^2+3)+12(2dl_3)^2((dml_3)^2+2)-5m^2(\beta l_0)^4\right)}{[(2dl_3)^2+(\beta l_0)^2]^{7/2}}K_1\left(z\right)
\Biggl\}.
\eea
}

Therefore, the Lorentz violation changes both the Casimir energy and pressure at finite temperature. The next step is to investigate this combined effect considering the space-like constant vector.

\subsubsection{Space-like constant vector}

Here, we have three possible cases associated with the space-like constant vector $n^\mu$ to analyze. Choosing $n^\mu=(0,1,0,0)$ or $n^\mu=(0,0,1,0)$, we find that the Casimir energy and Casimir pressure at finite temperature are given by Eqs. (\ref{LI00}) and (\ref{LI33}), respectively. Then there is no contribution due to Lorentz violation for this choice of the constant vector $n^\mu$.

It is important to note that if $n^\mu=(0,0,0,1)$ is chosen, the contribution associated with the Lorentz violation to the Casimir effect at finite temperature becomes imaginary. Therefore, since the Casimir effect is a real phenomenon, the space-like constant vector in the z-direction should be chosen as $n^\mu=(0,0,0,i)$. Then the Lorentz violation correction for the Casimir energy at finite temperature is given as
{\small
\bea
{\cal T}^{00(11)}_{LV}(\beta,d)&=&-\frac{2d g m}{\pi^2}\sum_{l_0,l_3=1}^{\infty}l_0\Biggl\{\frac{4m\left(-(\beta l_0)^2(3-(dml_3)^2)+(2dl_3)^2((mdl_3)^2+3)\right)}{[(2dl_3)^2+(\beta l_0)^2]^{3}}K_0\left(z\right)\nonumber\\
&+&\frac{\left(8(\beta l_0)^2((dml_3)^2-3)+4(2dl_3)^2(5(dml_3)^2+6)-3m^2(\beta l_0)^4\right)}{[(2dl_3)^2+(\beta l_0)^2]^{7/2}}K_1\left(z\right)
\Biggl\}
\eea
}
and for the Casimir pressure at finite temperature is
{\small
\bea
{\cal T}^{33(11)}_{LV}(\beta,d)&=&-\frac{4d g m}{\pi^2}\sum_{l_0,l_3=1}^{\infty}l_0\Biggl\{\frac{4m\left(-(\beta l_0)^2(3-(dml_3)^2)+(2dl_3)^2((mdl_3)^2+3)\right)}{[(2dl_3)^2+(\beta l_0)^2]^{3}}K_0\left(z\right)\nonumber\\
&+&\frac{\left(8(\beta l_0)^2((dml_3)^2-3)+4(2dl_3)^2(5(dml_3)^2+6)-3m^2(\beta l_0)^4\right)}{[(2dl_3)^2+(\beta l_0)^2]^{7/2}}K_1\left(z\right)
\Biggl\}.
\eea
}

In the very small mass limit, the Casimir energy at finite temperature with Lorentz violation corrections becomes
\bea
E(\beta,d)=-\frac{2}{\pi^2}\sum_{l_0,l_3=1}^\infty\left(\frac{(2dl_3)^2-3(\beta l_0)^2}{[(\beta l_0)^2+(2dl_3)^2]^3}+24gdl_0\frac{(2dl_3)^2-(\beta l_0)^2}{[(\beta l_0)^2+(2dl_3)^2]^4}\right)
\eea
and the Casimir pressure is
\bea
P(\beta,d)=-\frac{2}{\pi^2}\sum_{l_0,l_3=1}^\infty\left(\frac{3(2dl_3)^2-(\beta l_0)^2}{[(\beta l_0)^2+(2dl_3)^2]^3}+48gdl_0\frac{(2dl_3)^2-(\beta l_0)^2}{[(\beta l_0)^2+(2dl_3)^2]^4}\right),
\eea
where $E(\beta,d)\equiv {\cal T}^{00 (11)}(\beta,d)$ and $P(\beta,d)\equiv {\cal T}^{33 (11)}(\beta,d)$. Therefore, our results show that the Lorentz violating extension Myers-Pospelov in the scalar sector changes the Casimir effect at finite temperature.

\section{Conclusions}

Lorentz symmetry is fundamental to the Standard Model (SM) and General Relativity (GR). However, these theories are not complete. A fundamental theory must unify both into a single model that describes all interactions. Theories that seek this unification operate at very high energies, such as the Planck scale, where tiny violations of Lorentz symmetry emerge. To study these violations, which lead to new physics, effective theories have been constructed. One important effective theory developed to investigate Lorentz and CPT symmetries is the Standard Model Extension (SME), which encompasses all the physics of the SM and GR, along with additional Lorentz-violating operators. Here, we consider the scalar sector of the Myers and Pospelov model. In this model, Lorentz symmetry is broken by the introduction of a preferred four-vector $n^\mu$. The choice of this preferred four-vector yields different results in the investigation conducted in this work. To study certain applications arising from the Myers and Pospelov model, we utilize the TFD formalism. TFD possesses a rich topological structure that allows the analysis of different effects on an equal footing. Three different topologies are considered: (1) the topology $\Gamma_4^1=\mathbb{S}^1\times\mathbb{R}^{3}$, where the time-axis is compactified. This leads to the Stefan-Boltzmann law. (2) The topology $\Gamma_4^1$ with $\alpha=(0,0,0,i2d)$, where the compactification along the coordinate $z$ is considered. As a consequence, the Casimir effect at zero temperature is calculated. (3) The topology $\Gamma_4^2=\mathbb{S}^1\times\mathbb{S}^1\times\mathbb{R}^{2}$ where two axes are compactified, time and coordinate $z$. Then the Casimir effect at finite temperature is determined. It is important to note that our results demonstrate the dependence of the corrections to the Casimir effect and Stefan-Boltzmann law, associated with the massive scalar field in the presence of the Myers and Pospelov term, on the choice of the four-vector $n^\mu$. Different outcomes arise depending on whether this four-vector is timelike or spacelike.

\section*{Acknowledgments}

This work by A. F. S. is partially supported by National Council for Scientific and Technological Develo\-pment - CNPq project No. 312406/2023-1. L. H. A. R. F. thanks CAPES for financial support. The work of C.M.R has been partially supported by project Fondecyt Regular 1241369.


\global\long\def\link#1#2{\href{http://eudml.org/#1}{#2}}
 \global\long\def\doi#1#2{\href{http://dx.doi.org/#1}{#2}}
 \global\long\def\arXiv#1#2{\href{http://arxiv.org/abs/#1}{arXiv:#1 [#2]}}
 \global\long\def\arXivOld#1{\href{http://arxiv.org/abs/#1}{arXiv:#1}}


\begin{thebibliography}{100}

\bibitem{Kost1} V. A. Kostelecky and R. Potting, ``CPT, strings, and meson factories'',
\doi{10.1103/PhysRevD.51.3923} {Phys. Rev. D {\bf 51}, 3923 (1995).}

\bibitem{Kost2} V. A. Kostelecky and R. Potting, ``CPT and strings'',
\doi{10.1016/0550-3213(91)90071-5} {Nucl. Phys. B {\bf 359}, 545  (1991).}

\bibitem{Kost3} D. Colladay and V. A. Kostelecky, ``CPT violation and the standard model'',  
\doi{10.1103/PhysRevD.55.6760} {Phys. Rev. D {\bf 55}, 6760 (1997).}

\bibitem{Kost4} D. Colladay and V. A. Kostelecky, ``Lorentz violating extension of the standard model'',
\doi{10.1103/PhysRevD.58.116002} {Phys. Rev. D {\bf 58}, 116002 (1998).}

\bibitem{MP} R. C. Myers and M. Pospelov, ``Ultraviolet modifications of dispersion relations in effective field theory'',
\doi{10.1103/PhysRevLett.90.211601} {Phys. Rev. Lett. {\bf 90}, 211601(2003).}

\bibitem{Reyes1} L. Balart, C. M. Reyes, S. Ossandon and  C. Reyes, ``Perturbative unitarity and higher-order Lorentz symmetry breaking'',
\doi{10.1103/PhysRevD.98.035035} {Phys. Rev. D {\bf 98}, 035035  (2018).}

\bibitem{Pert} C. M. Reyes, L. Urrutia and J. D. Vergara, ``The scalar sector in the Myers-Pospelov model'', 
\doi{10.1063/1.2902786} {AIP Conf. Proc. {\bf 977}, 214 (2008).}

\bibitem{Umezawa1} Y. Takahashi and H. Umezawa, ``Thermo field dynamics'', 
\doi{ 10.1142/S0217979296000817} {Int. Jour. Mod. Phys. B {\bf 10}, 1755 (1996).}
 
\bibitem{Umezawa2} Y. Takahashi, H. Umezawa and H. Matsumoto, Thermofield Dynamics and Condensed States, North-Holland, Amsterdan, (1982).

\bibitem{Umezawa22} H. Umezawa, Advanced Field Theory: Micro, Macro and Thermal Physics, AIP, New York, (1993).

\bibitem{quart} F. C. Khanna, A. P. C. Malbouisson, J. M. C. Malboiusson and A. E. Santana, Themal quantum field theory: Algebraic aspects and applications, World Scientific, Singapore, (2009).

\bibitem{Khanna1} A. E. Santana and F. C. Khanna, ``Lie groups and thermal field theory'',
\doi{10.1016/0375-9601(95)00394-I} {Phys. Lett. A {\bf 203}, 68 (1995).}

\bibitem{Khanna2} A. E. Santana, F. C. Khanna, H. Chu, and C. Chang, ``Thermal Lie Groups, Classical Mechanics, and Thermofield Dynamics'',
\doi{10.1006/aphy.1996.0080} {Ann. Phys. {\bf 249}, 481 (1996).}

\bibitem{GBT} F. C. Khanna, A. P. C Malbouisson, J. M. C. Malbouisson and A. E. Santana, ``Quantum fields in toroidal topology,''
\doi{10.1016/j.aop.2011.07.005} {Ann. Phys. {\bf 326}, 2634 (2011)}.

\bibitem{Moeller} N. Moeller and B. Zwiebach, ``Dynamics with infinitely many times derivatives and rolling tachyons'', 
\doi{10.1088/1126-6708/2002/10/034} {JHEP {\bf 10}, 034 (2002).} 

\bibitem{Cheng:2002rz}
T.~C.~Cheng, P.~M.~Ho and M.~C.~Yeh, ``Perturbative approach to higher derivative and nonlocal theories,''
\doi{10.1016/S0550-3213(02)00020-2} {Nucl. Phys. B \textbf{625}, 151-165 (2002)}, \arXiv{hep-th/0111160}{hep-th}.
T.~C.~Cheng, P.~M.~Ho and M.~C.~Yeh, ``Perturbative approach to higher derivative theories with fermions,''
\doi{10.1103/PhysRevD.66.085015} {Phys. Rev. D \textbf{66}, 085015 (2002)}, \arXiv{hep-th/0206077}{hep-th}.




\end{thebibliography}
\end{document}